\documentstyle[12pt,epsfig]{article}
\begin{document}

\begin{titlepage}
\rightline{August 2012}
\vskip 2cm
\centerline{\Large \bf
Implications of mirror dark matter kinetic mixing}
\vskip 0.4cm
\centerline{\Large \bf
 for CMB anisotropies}

\vskip 2.2cm
\centerline{R. Foot\footnote{
E-mail address: rfoot@unimelb.edu.au}}

\vskip 0.7cm
\centerline{\it ARC Centre of Excellence for Particle Physics at the Terascale,}
\centerline{\it School of Physics, University of Melbourne,}
\centerline{\it Victoria 3010 Australia}
\vskip 2cm
\noindent
Mirror dark matter is a dissipative and self-interacting multiparticle dark matter 
candidate which can explain the DAMA, CoGeNT and CRESST-II direct detection experiments.
This explanation requires photon-mirror photon kinetic mixing of strength $\epsilon \sim 10^{-9}$.
Mirror dark matter with such kinetic mixing can potentially leave distinctive 
signatures on the CMB anisotropy spectrum. 
We show that the most important effect of kinetic mixing on the 
CMB anisotropies is the suppression of the height of the third and higher odd peaks. 
If $\epsilon \stackrel{>}{\sim} 10^{-9}$ then this feature can
be observed by the PLANCK mission in the near future.

 \end{titlepage}

A large variety of observations have lead to a simple picture of the Universe.
In a nutshell, we live in a spatially flat, expanding Universe consisting of dark energy ($\sim 70\%$), 
non-baryonic dark matter ($\sim 26\%$) and ordinary baryons
($\sim 4\%$).  
In the last two decades, detailed observations of the cosmic microwave background (CMB) 
by COBE\cite{cobe}, WMAP\cite{wmap}, SPT\cite{spt} and 
many other missions have provided important tests of this basic picture. 
A key question concerns the identity of non-baryonic dark matter.
Although it is popular to assume that dark matter consists of a single species of weakly 
interacting massive particles (standard cold dark matter model),
in actuality the particle physics underlying the non-baryonic dark matter content in the 
Universe is currently unknown, as is the physics responsible for the dark energy.

One thing we do know, though, is that
the standard model has been very successful in describing the interactions of the ordinary particles.
In fact, it is possible that
such a structure might also be responsible for the non-baryonic 
dark matter in the Universe as well. That is,
dark matter might consist of a hidden (mirror) sector with particles and interactions 
exactly isomorphic to the 
ordinary ones\cite{ly, flv} (for a review, see e.g.\cite{review}).  
Provided that initially the mirror sector temperature is much less than in the ordinary 
sector, i.e. $T' \ll T$ in the early Universe, such a scenario
can explain the large scale structure of the Universe in a way completely analogous to 
standard cold dark matter\cite{lss,paolo}. 

On much smaller scales, mirror dark matter is radically different to standard cold dark matter.
It is self-interacting, dissipative and 
multi-component\footnote{
There are potential limits on self interactions of dark matter from 
observations of the Bullet cluster\cite{bullet}.
These observations can
set limits on dark matter self interactions provided that the bulk
of the dark matter particles are distributed throughout the
cluster and not bound to individual galaxies. 
However mirror dark matter is dissipative
and in clusters (or at least in some of them)
the bulk of the dark matter particles might be confined in galactic
halos [cf. ref.\cite{zurab}]. 
Under this assumption mirror
dark matter is consistent with Bullet cluster observations.
Precisely how much mirror dark matter can be in the form of intergalactic gas requires further study.
}.
These properties might help explain some puzzling aspects of dark matter
on small scales, such as the inferred cored central density 
profiles in galaxies [cf. ref.\cite{spergel}].
At the current epoch dark matter needs to be roughly spherically distributed
in spiral galaxies to be consistent with various observations
\footnote{
Deviations from perfectly spherical halos are also required and might be due
to various sources including, e.g. partial collapse of the halo due to dissipative
processes, mirror magnetic fields, asymmetric heating from ordinary supernovae which are
distributed in the disk (i.e. they are not spherically
distributed), possible existence of a dark disk subcomponent etc. }.
However, within galaxies mirror dark 
matter would be expected to collapse into a disk, analogous to the
way in which ordinary matter collapses, unless a significant heat source exists. 
Ordinary
supernovae can potentially supply the required energy provided that photon-mirror photon
kinetic mixing\cite{he,holdom}, 
\begin{eqnarray}
{\cal L}_{mix} = \frac{\epsilon}{2} F^{\mu \nu} F'_{\mu \nu}
\label{kine}
\end{eqnarray}
of strength $\epsilon \sim 10^{-9}$ exists\cite{sph}. 
[In the above equation, $F_{\mu \nu}$ ($F'_{\mu \nu}$) is the ordinary (mirror) photon
field strength tensor.]  If $\epsilon \sim 10^{-9}$ then ordinary core collapse supernovae would release their energy into
$e'^\pm, \ \gamma'$ in roughly equal proportion to neutrinos\cite{raffelt}.
The produced $e'^\pm$ can scatter off halo $e'$, while mirror photons with
energies up to 10's of keV can be absorbed by the halo if there is a substantial metal (e.g. $Fe'$) component 
due to the large photoionization cross-section for heavy elements\cite{sph}. 
Thus, it seems possible that a significant
fraction of the energy of core collapse supernovae can be transferred to the halo.
The net effect is that the ordinary and dark matter components of galaxies might be in a kind of
dynamical equilibrium where the energy supplied to the halo by ordinary supernovae balances the
energy lost to the halo due to radiative cooling.
Since the former is related to the   
galactic luminosity and the latter the dark matter density, that is, $v_{rot}$,
it has been speculated\cite{foots} that
this might potentially explain puzzling regularities on small scales, such as
the Tully-Fisher relation\cite{tf}. 

Photon-mirror photon kinetic mixing of strength $\epsilon \sim 10^{-9}$ is also implicated\cite{footearly}
by the positive results of the direct detection experiments, DAMA\cite{dama}, CoGeNT\cite{cogent} and 
CRESST-II\cite{cresst}.
It has been shown in ref.\cite{foot1} that these experiments can be explained\footnote{
This explanation has some tension with the null result of XENON100\cite{xenon100}. 
The amount of tension depends on the recoil energy threshold assigned for XENON100.  Updating
the earlier estimate of ref.\cite{foot1}, using the latest XENON100 results\cite{xenon100}, we 
find that consistency of this mirror dark matter explanation requires a XENON100 energy threshold of
13 keV; around a factor of two higher than the XENON100 estimate.
Nevertheless, the XENON100 energy scale is highly uncertain and even a factor of two 
uncertainty in their threshold energy appears to be possible\cite{damcol}.}
by the interactions 
of a halo $Fe'$ component with $\epsilon \sqrt{\xi_{Fe'}} \approx 2\times 10^{-10}$.
Here $\xi_{Fe'}$ is the abundance by mass of the halo $Fe'$ component (at the Earth's
location) normalized to $0.3\ GeV/cm^3$.
Other parameter space is also possible.  
Naturally, it is very difficult to predict $\xi_{Fe'}$
with any certainty, but of course we expect $\xi_{Fe'} \stackrel{<}{\sim} 1$, and thus $\epsilon 
\stackrel{>}{\sim} 2\times 10^{-10}$.

Kinetic mixing can also have important implications for cosmology. 
Successful big bang nucleosynthesis (BBN) and large scale structure (LSS) 
require the initial condition $\rho_{\gamma'} \ll \rho_\gamma$ and $n_{b'} \approx 5n_b$. 
How such initial conditions might have arisen has been discussed in the literature\cite{discussed}.
However, if photon-mirror photon kinetic mixing exists then this will generate entropy in the
mirror sector via 
the process $e\bar e \to e'\bar e'$ when $T_\gamma \stackrel{>}{\sim} m_e$\cite{cg}.
That is
$T'_\gamma$ will be generated
even if we start with $T'_\gamma \ll T_\gamma$. 
In fact, it has been shown that
the asymptotic value of the ratio: $T'_\gamma/T_\gamma$, which we here define as $x$, is given by\cite{p1,f1}
\begin{eqnarray}
x  \simeq 0.31 \left( {\epsilon \over 10^{-9}}\right)^{1/2}\ .
\label{1}
\end{eqnarray}
The value $x \sim 0.3$ is close to the limit estimated from the matter power spectrum
i.e. successful large scale structure\cite{paolo}. 
We will estimate that the upper bound 
on $x$ from such considerations is conservatively around $x \stackrel{<}{\sim} 0.5$.
Non-zero $T'_\gamma/T_\gamma$ will also lead to important effects for the CMB as previously
discussed in ref.\cite{paolo} (see also ref.\cite{lss}).
In view of the forthcoming results from the PLANCK mission it is pertinent to 
examine thoroughly the possible effects that 
kinetic mixing will induce for the CMB. This is the purpose of this paper.
In fact, we will show that mirror dark matter can potentially leave a distinctive imprint
on the tail of the CMB anisotropy spectrum. Our most important observation is that the
height of the third and higher odd peaks can be suppressed. This should be 
observable by PLANCK provided
$\epsilon \stackrel{>}{\sim} 10^{-9}$.

To summarize, we assume a mirror sector exactly isomorphic to the ordinary one, except with initial
conditions $T'_\gamma \ll T_\gamma$. Ordinary and mirror particles can interact via photon - mirror photon
kinetic mixing, which can excite the mirror degrees of freedom in the early Universe. 
In particular the process:
$e\bar e \to e' \bar e'$
generates the mirror particles until the $e\bar e$ have annihilated at $T_\gamma \sim m_e$, 
the final $T'_\gamma/T_\gamma$ value given in Eq.(\ref{1}). 
In fact, most of the mirror entropy generation occurs after the neutrinos have decoupled.
One effect of this is to induce a slight cooling of the ordinary photons relative to the ordinary neutrinos.
The net effect is that there is additional neutrino energy density and also an
additional relativistic component comprised of mirror photons. These two additional components 
to the relativistic energy density can be parameterized
in terms of extra neutrino degrees of freedom\cite{f1}:
\begin{eqnarray}
\delta N_{eff}^a (\epsilon) [CMB] &=& 3 \left( \left[
{T_\nu (\epsilon) \over T_\nu (\epsilon = 0)}\right]^4\ - \ 1\right)
\nonumber \\
\delta N_{eff}^b (\epsilon) [CMB] &=& \  {8 \over 7}\left( {T'_\gamma (\epsilon) \over T_\nu (\epsilon =0)}
\right)^4  \ .
\label{add}
\end{eqnarray}
Here, the temperatures are evaluated at the time when photon decoupling occurs, i.e. when 
$T_\gamma = T_{dec} \approx 0.26$ eV. Using the result from Eq.(\ref{1}), together with the usual $T_\nu/T_\gamma$ relation,
we have 
\begin{eqnarray}
\delta N_{eff}^b (\epsilon)[CMB]
&\simeq & {8x^4 \over 7}\left({11 \over 4}\right)^{4/3}
\simeq  0.041 \left({\epsilon \over 10^{-9}}\right)^2\ .
\end{eqnarray} 
Also, numerical work\cite{f1} has found that $\delta N_{eff}^a (\epsilon)[CMB] \approx 0.8 N_{eff}^b(\epsilon)[CMB]$.
This additional energy density can directly affect the predicted CMB anisotropies.
In fact, it is known that additional relativistic energy density can dampen the CMB tail\cite{bsaj,dampen}.
However, there is another important effect for the CMB.
If dark matter consists of mirror particles, then they experience significant pressure
prior to mirror photon decoupling. 
If $T'_\gamma < T_\gamma$ then this epoch occurs prior to the familiar hydrogen recombination. 
One can anticipate that the small scale inhomogeneities in the mirror matter density should
be suppressed, since the Fourier modes which enter the horizon before the time of mirror hydrogen
recombination undergo acoustic oscillations due to the pressure of the mirror baryon-photon fluid
\footnote{
The mirror photons can undergo diffusion (Silk damping). This would washout small scale inhomogeneities in the
mirror radiation field just before mirror photon decoupling.
However it should have very little effect on the ordinary CMB or matter power
spectrum since $\rho_{\gamma'}$ is a very small component to the overall energy
density (assuming $x\stackrel{<}{\sim} 0.5$). Of course 
mirror photon diffusion would be expected to significantly damp the tail of the 
mirror CMB anisotropies, just like ordinary photon diffusion does for the ordinary CMB. 
}.
In other words, we expect a suppression of power on small scales when compared with
standard non-interacting cold dark matter. The previous study\cite{paolo} has indeed observed this
effect on the matter power spectrum.
This suppression of power on small scales will also influence the CMB spectrum, and one would
anticipate that this might also dampen the CMB anisotropies at high multipoles.
This effect, is of course, in addition to the effect of the increased relativistic energy density
due to $\delta N_{eff}^a + \delta N_{eff}^b$. [Both effects will be included in our numerical work.]
It turns out that the mirror baryon acoustic oscillation effect is not only larger in magnitude, but has
the distinctive feature of suppressing the higher odd peaks more than the even ones.

Although the effect of additional relativistic energy density has been well studied in the literature (see e.g.
ref.\cite{bsaj, dampen}),
and can be explored using existing CMB codes, the mirror baryon acoustic oscillation 
effect on the mirror dark matter perturbations requires modifications. The relevant equations, though,
are a straightforward generalization to the equations governing the perturbations of the ordinary 
baryons and photons.
Our strategy is to numerically solve these equations, essentially using
techniques developed in refs.\cite{cmbfast} (see also \cite{sjak,hu}).
A very clear and helpful review is the one given by Dodelson\cite{dodelson}. 
 
Recall, the anisotropy spectrum today is characterized in terms of $C_{\ell}$. These quantities  
are the variance of the
coefficients, $a_{\ell m}$ in the expansion of the photon temperature field 
in terms of spherical harmonics, $Y_{\ell m}$, i.e.
\begin{eqnarray}
\langle a_{\ell m} a^*_{\ell' m'} \rangle = \delta_{\ell \ell'} \delta_{m m'} C_\ell\ .
\end{eqnarray}
The terms $C_\ell$ can be related to the $\ell^{th}$ multipole moment, $\Theta_\ell$, 
in the Legendre expansion
of the Fourier transformed photon temperature field via the equation:
\begin{eqnarray}
C_\ell = {2 \over \pi} \int^{\infty}_0 dk k^2 P_{\Phi}^i |\Theta_\ell (k,\eta_0)/\Phi^i|^2
\ 
\label{cell}
\end{eqnarray}
where $P_{\Phi}^i$ is the initial power spectrum of the metric perturbation with
initial value $\Phi^i$.
Finally the moments $\Theta_\ell (k,\eta_0)$ today 
can be related to the perturbations $\Theta_0 (k, \eta)$,  $v_b (k, \eta)$,  
$\Phi (k, \eta)$,  $\Psi (k, \eta)$, $\Pi (k, \eta)$ 
near photon decoupling [where $v_b (k, \eta)$, $\Psi (k, \eta)$, $\Phi (k, \eta)$, 
$\Pi (k, \eta)$ are the baryonic velocity,
metric perturbations, and polarization tensor respectively].
The critical equation is\cite{dodelson}:
\begin{eqnarray}
\Theta_{\ell} (k, \eta_0) = \int^{\eta_0}_0 d\eta g(\eta)
\left( \Theta_0 + \Psi + {1 \over 4}\Pi + {3 \over 4k^2}{d^2 \over d\eta^2} 
[g(\eta) \Pi]\right)  j_\ell [k(\eta_0 - \eta)] \nonumber \\
+ \int^{\eta_0}_0 d\eta g(\eta)
iv_b \left( j_{\ell-1} [k(\eta_0 - \eta)] - {(\ell +1)j_\ell [k(\eta_0 - \eta)] \over 
k(\eta_0 - \eta)} \right) \nonumber \\
+ \int^{\eta_0}_0 d\eta \ e^{-\tau} \left[ \dot{\Psi} - 
\dot{\Phi}\right] j_\ell [k(\eta_0- \eta)]\ 
\label{cell2}
\end{eqnarray}
where $\tau$ is the optical depth for Thomson scattering and 
$g(\eta) \equiv - \dot{\tau} e^{-\tau}$ is the visibility function, which peaks near
photon decoupling.
The evolution of the quantities $\Theta_0, v_b$, $\Psi$, $\Phi$ and $\Pi$ 
are governed by a set of linear
equations,
arising from the Boltzmann-Einstein equations. We assume standard 
adiabatic scalar initial conditions.
The relevant equations are given in the appendix.

It is important to note that the mirror dark matter model introduces only one additional parameter,
$x \equiv T'_\gamma/T_\gamma$ 
which is related to the fundamental Lagrangian kinetic mixing parameter, $\epsilon$ via Eq.(\ref{1}).
The cosmological evolution of
mirror dark matter, in the limit where $x \to 0$ (i.e. $\epsilon \to 0$) exactly mimics cold dark matter.
This is because mirror particles feel
negligible pressure after the mirror photon decoupling 
epoch, $t'_{dec}$, and $t'_{dec} \to 0$ as $x \to 0$.
As $x$ increases from zero, differences begin to appear.  Our job now is to determine 
what the observable differences are.
To study these effects for the CMB one cannot simply choose a particular point for
the parameters $\Omega_m h^2, \Omega_b h^2, h, ...$ from a fit assuming standard cold dark matter
and vary $x$. 
Doing this, for example, would modify the epoch of
matter radiation equality, $z_{EQ} + 1  = \Omega_m/\Omega_r$, due 
to the additional contributions [Eq.(\ref{add})]
to $\Omega_r$. The matter radiation equality has been precisely
constrained by the data and thus any modification to $z_{EQ}$ by new physics
needs to be compensated for by adjustments to the parameters (in this case, $\Omega_m h^2$).
In fact, what needs to be done is to examine parameter space where not only
$z_{EQ}$ is fixed, but also $\Omega_b h^2$ and $\theta_s$ (the angular size of the sound
horizon at decoupling), since these quantities
have also been precisely determined by the data. A similar situation has been noted when
considering the effect of additional relativistic neutrino degrees of freedom\cite{dampen, bsaj}.
In this parameter space direction, the observable effects from
varying $x$ occur at small angular scales.

It is reasonably straightforward to write a code to numerically solve the relevant set of equations
to obtain the CMB anisotropy spectrum.
For a given set of parameters, $\Omega_b h^2, \Omega_m h^2, h,  ...$, 
comparison of our code with existing high accuracy codes, e.g. CMBFAST\cite{nasa}, confirms that
our computation of the $C_\ell$ values are accurate to within a few percent. 
This is sufficient for making a comparison of mirror dark matter with standard cold dark matter.

In figure 1,2,3 we give our results for the CMB spectrum.  We consider
a flat Universe with the reference parameters 
$\Omega_m h^2 = 0.14$, $\Omega_b h^2 = 0.022$, $\Omega_{\Lambda} = 1 - \Omega_m$, $h = 0.70$
[$\Omega_m \equiv \Omega_b + \Omega_{b'}$]. 
These reference parameters are defined at $x = 0$.
As discussed above, these parameters are adjusted as $x$ is
varied such that
$z_{EQ}$, $\Omega_b h^2$ and $\theta_s$ are held fixed.
[We also adjust the overall normalization by fixing the height of the first peak.]
A scale invariant initial perturbation spectrum 
(Harrison-Zel'dovich and Peebles spectrum) is assumed and 
we have neglected reionization effect. 
Since we are interested in comparing mirror dark matter versus standard non-interacting
cold dark matter (cosmologically equivalent to mirror dark matter with $x = 0$) 
small effects due to primordial tilt or reionization are not important to leading order.
Figure 1 illustrates the expected
agreement at large angular scales, as we vary $x$.
In figure 2, we consider the small angular scale region of interest.
In figure 3 we plot $F_\ell (x) \equiv C_\ell (x)/C_\ell (x = 0)$ for several values of $x$.

\centerline{\epsfig{file=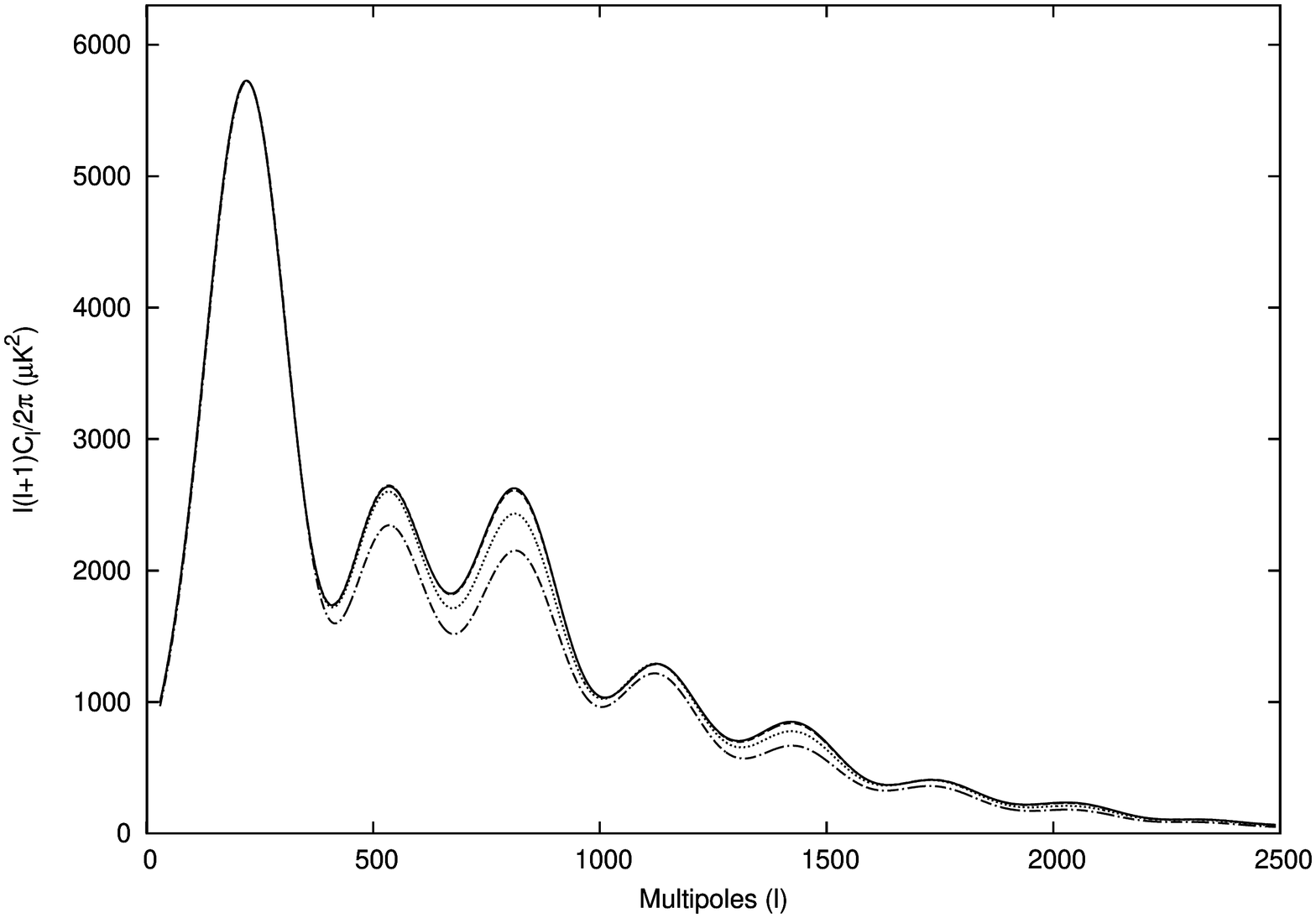,angle=270,width=13.2cm}}
\vskip 0.2cm
\noindent
{\small
Figure 1: The anisotropy spectrum for mirror dark matter versus standard cold dark matter. 
The solid line is standard cold dark matter model with parameters described in the text
(equivalent to mirror dark matter with $x=0$), while
mirror dark matter with $x=0.3$ (dashed line), $x = 0.5$ (dotted line) and $x=0.7$ (dashed-dotted line) 
are also shown.  
}

\vskip 0.1cm

\centerline{\epsfig{file=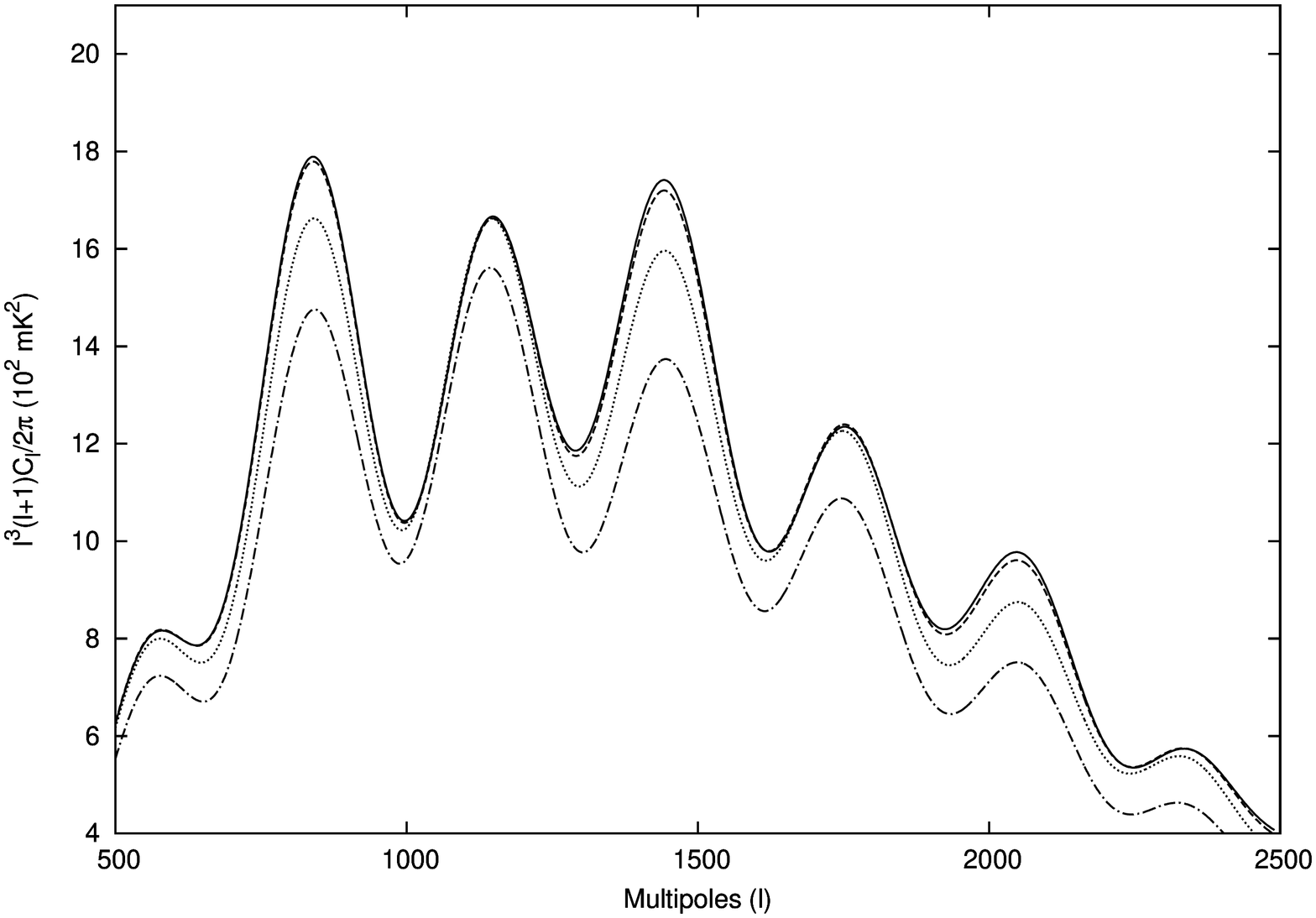,angle=270,width=13.2cm}}
\vskip 0.2cm
\noindent
{\small
Figure 2: The CMB tail. The curves correspond to the same parameters as figure 1.
}

\vskip 0.3cm


\vskip 0.3cm

\centerline{\epsfig{file=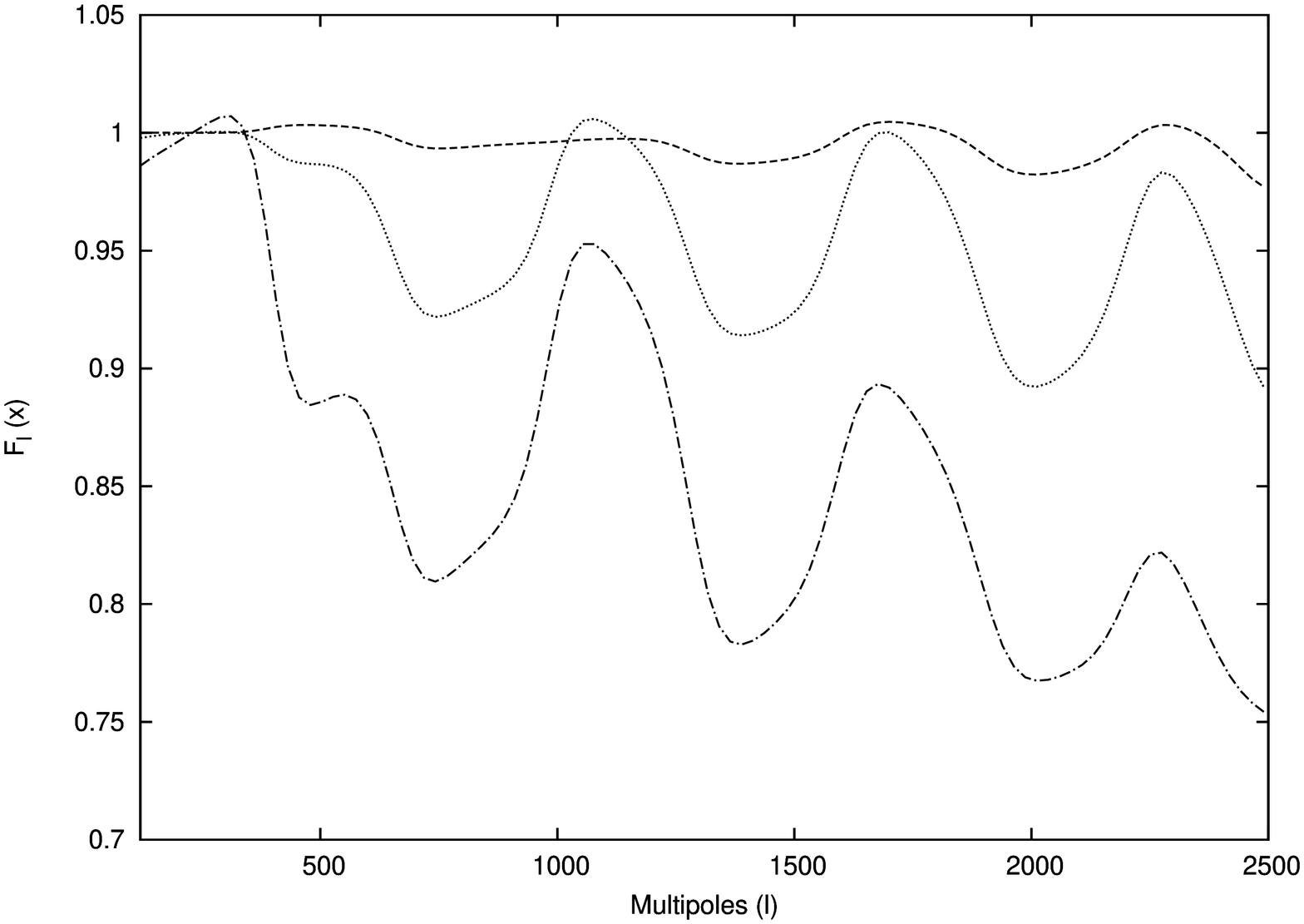,angle=270,width=13.2cm}}
\vskip 0.3cm
\noindent
{\small
Figure 3: $F_\ell (x) \equiv C_{\ell}(x)/C_{\ell} (x = 0)$ for $x = 0.3$ (dashed line) and
$x = 0.5$ (dotted line), and $x = 0.7$ (dash-dotted line) are shown.
}

\vskip 0.8cm

Figures 2,3 clearly show the expected suppression of anisotropies at small angular scales, starting
around the third peak. Interestingly,
we see that the suppression is larger for the higher odd peaks than the even ones.
These features can be readily understood. Odd peaks arise from compressions of the baryon-photon
fluid, even peaks are rarefactions. 
When the gravitational driving force is suppressed, one expects the
odds peaks (the compressions) to be more affected than the even peaks
(related effects occur when $\Omega_b h^2$ is reduced). Furthermore, the differences only become
apparent for the higher peaks because the suppression of power only occurs at small scales.

Currently the most accurate measurement of the CMB damping tail has been made with the
South Pole telescope\cite{spt}. These measurements show a slight damping, 
around $\sim 2.5\%$ at $\ell \sim 2000$ cf. predictions
of the standard cold dark matter model. 
This damping provides an interesting hint that 
$x \approx 0.4$ [i.e. $\epsilon \approx 2\times 10^{-9}$ from Eq.(\ref{1})].
In any case,
these observations limit $x \stackrel{<}{\sim} 0.5$ [or $\epsilon 
\stackrel{<}{\sim} 3\times 10^{-9}$]. It is anticipated that the PLANCK mission should 
improve the precision, which will probe $\epsilon$ in the range: $10^{-9} 
\stackrel{<}{\sim} \epsilon \stackrel{<}{\sim} 3 \times 10^{-9}$.

In addition to CMB anisotropies the matter power spectrum can also be used
to constrain parameters. However since small scales $k \stackrel{>}{\sim} 0.1\ h\ Mpc^{-1}$
have gone nonlinear today, we consider the matter power spectrum
on larger scales than this [linear regime]. 
It is straightforward to compute the power spectrum of matter,
\begin{eqnarray}
P(k) = 2\pi^2 \delta_H^2 {k \over H_0^4} T^2(k)
\end{eqnarray}
where $H_0 = 100h\ km \ sec^{-1} Mpc^{-1}$ is the Hubble rate today and
$T(k)$ is the transfer function (see e.g. ref.\cite{dodelson} for details).
In figure 4 we compare the obtained matter power spectrum
for the various $x$ values considered, for the same parameters used in figures 1-3.
[Recall, $\Omega_m, \Omega_b$ and $h$ are varied as $x$ changes such 
that $z_{EQ}, \ \Omega_b h^2$ and $\theta_s$ are fixed.]
As expected, deviations only occur on small scales as $x$ increases from zero.
This figure indicates that a rough bound of $x \stackrel{<}{\sim} 0.3-0.5$ could
be extracted from galaxy surveys. Also note that very similar results to our figure 4 
have been obtained in the earlier study\cite{paolo}.

\vskip 0.2cm
\centerline{\epsfig{file=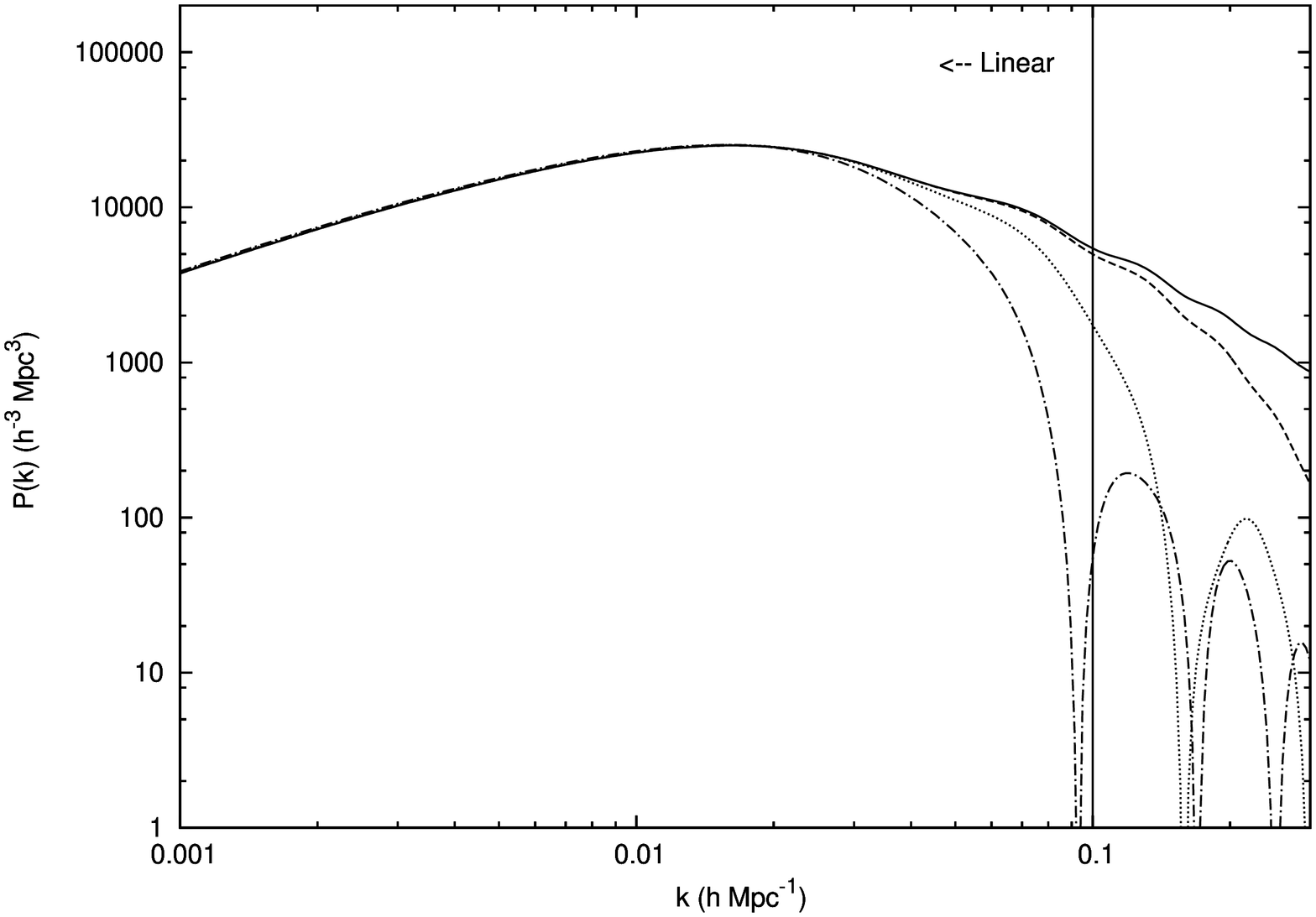,angle=270,width=13.2cm}}
\vskip 0.2cm
\noindent
{\small
Figure 4: Power spectrum of matter for the same parameters as figure 1. As in figure 1, 
$x = 0$ (solid line), $x = 0.3$ (dashed line), $x=0.5$ (dotted line) and
$x = 0.7$ (dashed-dotted line).
}

\vskip 1.0cm

In conclusion,
we have examined the implications of kinetically mixed mirror dark 
matter for CMB anisotropies. 
This dark matter candidate can potentially
leave distinctive signatures on the CMB spectrum. We have found that the most
important effects of kinetic mixing on CMB anisotropies is the suppression of
the height of the third and higher odd peaks. This effect will be
sensitively probed by the PLANCK mission in the near future.

\vskip 1.0cm
\noindent
{\bf Appendix - Linear perturbation theory with mirror dark matter}
\vskip 0.4cm

The relevant equations governing the linear evolution of scalar perturbations in the Universe 
has a rich history starting with the work of Lifshitz in 1946\cite{shitz} and developed by 
many others, e.g. ref.\cite{others}. For an up to date review see ref.\cite{dodelson}.
As summarized in that review, the relevant equations governing the moments of the photon
distribution (including the polarization field), which we consider numerically up to 
order $\ell = 5$
together with corresponding moments for
neutrinos and baryonic matter perturbations, in the conformal Newtonian gauge, 
are:
\begin{eqnarray}
\dot{\Theta}_0 + k\Theta_1 &=& -\dot{\Phi} \nonumber \\
\dot{\Theta}_1 - {k \over 3}\Theta_0 + {2k \over 3} \Theta_2 &=& {k \over 3}\Psi +
\dot{\tau}\left[ \Theta_1 - {iv_b \over 3}\right]
\nonumber \\
\dot{\Theta}_{\ell} - {k\ell \over 2\ell + 1}\Theta_{\ell -1}  + {k(\ell+1) \over 2\ell + 1} \Theta_{\ell+1} 
&=& \dot{\tau} \left[\Theta_{\ell} - \delta_{\ell 2} {\Pi \over 10}\right]\ , \ \ell \ge 2
\nonumber \\
\Pi &=& \Theta_2 + \Theta_{P2} + \Theta_{P0} \nonumber \\
\dot{N_0} + kN_1 &=& -\dot{\Phi} \nonumber \\
\dot{N_1} - {k \over 3}N_0 + {2k \over 3} N_2 &=& {k \over 3}\Psi 
\nonumber \\
\dot{N_\ell} - {k\ell \over 2\ell + 1}N_{\ell - 1} + {k(\ell + 1) \over 2\ell + 1} N_{\ell + 1} &=& 0 \ , \ \ell \ge 2 
\nonumber \\
\dot{\delta_b} + ikv_b &=& -3\dot{\Phi}
\nonumber \\
\dot{v_{b}} + {\dot{a} \over a} v_{b} &=& -ik\Psi 
+ {\dot{\tau} \over R} \left[ v_b + 3i\Theta_1 \right] \nonumber \\
\dot{\Theta}_{P0} + k\Theta_{P1} &=& \dot{\tau} \left[ \Theta_{P0} - {\Pi \over 2}\right] \nonumber \\
\dot{\Theta}_{P\ell} - {k\ell \over 2\ell + 1}\Theta_{P(\ell - 1)} + {k(\ell + 1) \over 2\ell + 1}\Theta_{P(\ell+1)}
&=& \dot{\tau} \left[ \Theta_{P\ell} - \delta_{\ell 2}{\Pi \over 10}\right]\  , \ \ell \ge 1
\label{9x}
\end{eqnarray}
where $\dot{\tau}  \equiv -X_e (1-Y_{p}) n_b \sigma_T a$,
$Y_{p} \simeq 0.24$ is the primordial helium mass fraction, 
$\sigma_T$ is the Thomson cross-section and 
$R \equiv {3\rho_b \over 4\rho_\gamma}$.
For the mirror sector, we have an analogous set of equations with $\Theta_{\ell} \to \Theta'_{\ell}$,
$\Theta_{P\ell} \to \Theta'_{P\ell}$
$N_{\ell} \to N'_{\ell}$ ($\ell \ge 0$),
$\delta_b \to \delta'_b$, $v_b \to v'_b$ and $\dot{\tau}, R \to \dot{\tau}', R'$.
Here $\dot{\tau'} \equiv -X_{e'} (1-Y'_{p}) n_{b'} \sigma_T a$ and $R' \equiv {3\rho_{b'} 
\over 4\rho_{\gamma'}}$.
Compared with the standard cold dark matter model,
the only additional parameter introduced is
$x \equiv T'_\gamma/T_\gamma$ which is related to $\epsilon$ via Eq.(\ref{1}).
[Recall these equations reduce to the equations governing standard 
cold dark matter when $x \to 0$ and
$\rho_{b'} \to \rho_c$].
Finally, we have the two relevant Einstein equations:
\begin{eqnarray}
k^2(\Phi + \Psi) &=& -32\pi G a^2 (\rho_\gamma \Theta_2 + \rho_\nu N_2 +
\rho_{\nu'} N'_2 + \rho_{\gamma'} \Theta'_2)
\nonumber \\
k^2 \Phi + 3{\dot{a} \over a}\left( \dot{\Phi} - \Psi {\dot{a} \over a}
\right) &=& 4\pi G a^2 [\rho_b \delta_b + \rho_{b'} \delta_{b'} + 
4\rho_\gamma\Theta_0  + 4\rho_\nu N_0 + 4\rho_{\gamma'}\Theta'_0 + 4\rho_{\nu'} N'_0] \ .
\nonumber \\
.
\label{10x}
\end{eqnarray}
where $\rho_{\gamma'} = x^4 \rho_\gamma$ and $\rho_\nu = N_{eff} (7/8)(4/11)^{4/3}\rho_\gamma$,
$N_{eff} = 3.046 + \delta N_{eff}^a$.
For our application we can neglect $N'_\ell, \rho_{\nu'}$ because we have negligible excitation of the mirror neutrino
degrees of freedom.
All derivatives in Eqs.(\ref{9x},\ref{10x}) are with respect to conformal time, $\eta$.
The quantity, $X_e$ is the free electron fraction
[$X_e \equiv n_e/n_H$ where $n_H$ is the total number of hydrogen nuclei].
It obeys the Boltzmann equation\cite{peebles,dodelson} 
\begin{eqnarray}
{1 \over a}{dX_e \over d\eta} = \left[ (1-X_e)\beta - X_e^2 (1-Y_{p})n_b \alpha^{(2)}\right] C
\end{eqnarray}
where 
\begin{eqnarray}
\beta &=& \langle \sigma v \rangle \left( {m_e T_\gamma \over 2\pi}\right)^{3/2} \ e^{-\epsilon_0/T_\gamma}
\nonumber \\
\alpha^{(2)} &=& \langle \sigma v \rangle \simeq 
9.78 {\alpha^2 \over m_e^2} \left( {\epsilon_0 \over T_\gamma}\right)^{1/2} \ ln \left(
{\epsilon_0 \over  T_\gamma}\right)\  \nonumber \\
C &=& {\Lambda_\alpha + \Lambda_{2\gamma} \over 
\Lambda_\alpha + \Lambda_{2\gamma} + \beta e^{3\epsilon_0/4T_\gamma}} \ .
\end{eqnarray}
Here $\epsilon_0 = 13.6$ eV is the binding energy of Hydrogen, 
$\Lambda_{2\gamma} = 8.227 \ sec^{-1}$ and
$\Lambda_\alpha = H(3\epsilon_0)^3/[(8\pi)^2 (1 - X_e)n_b (1 - Y_{p})]$. 
A similar set of equations will govern $X_{e'}$ (with $Y_{p} \to Y'_{p},\ T_\gamma \to T'_\gamma,\ n_b \to n_{b'}$).  
Evidently, the latter depends on the primordial mirror helium
mass fraction, $Y'_{p}$. This quantity can be computed solving the relevant mirror BBN
equations, and for $\epsilon \sim 10^{-9}$, is\cite{fp2} $Y'_{p} \approx 0.85$. 
[Note that $Y'_{p}$ is a slowly varying function of $\epsilon$ if
$\epsilon \sim 10^{-9}$ .]

The equations must be supplemented with initial conditions. We consider the standard adiabatic
scalar perturbations. 
We further assume that the
initial perturbation of $\Phi$ is drawn from a scale invariant Gaussian distribution
with mean zero and variance, parameterized in the usual way: 
$P^i_\Phi = (50\pi^2/9k^3)\delta_H^2 (\Omega_m/D_1(a=1))^2$.

The above set of equations, together with the Friedmann equation are numerically solved for $k$
values on a logarithmically spaced grid between $[k_{min}, \ k_{max}]$. For our
numerical work $k_{min} = 20/\eta_0$ and $k_{max} = 6000/\eta_0$. The $C_\ell$ values are 
then obtained from Eqs.(\ref{cell},\ref{cell2}).

\vskip 0.3cm
\noindent
{\bf Acknowledgments}

\vskip 0.1cm
\noindent
This work was supported by the Australian Research Council.

\end{document}